\documentstyle[12pt]{article}
\newcommand{\beq}{\begin{equation}}
\newcommand{\eeq}{\end{equation}}
\newcommand{\eq}[1]{(\ref{#1})}
\newcommand{\A}{{\bf a}}
\newcommand{\B}{{\bf b}}
\newcommand{\C}{{\bf c}}
\newcommand{\beqn}{\begin{eqnarray}}
\newcommand{\eeqn}{\end{eqnarray}}
\newcommand{\dst}{&\displaystyle}
\newcommand{\al}{\mbox{$Z\alpha$}}
\newcommand{\eps}{\mbox{$\varepsilon$}}
\newcommand{\Q}{\mbox{$\kappa$}}
\newcommand{\r}{\mbox{${\bf r}$}}
\newcommand{\k}{\mbox{${\bf k}$}}
\newcommand{\vt}{\mbox{\boldmath ${\theta}$\unboldmath}}
\newcommand{\g}{\mbox{\boldmath ${\gamma}$\unboldmath}}
\newcommand{\ro}{\mbox{\boldmath ${\rho}$\unboldmath}}
\newcommand{\val}{\mbox{\boldmath ${\alpha}$\unboldmath}}
\newcommand{\vl}{\mbox{\boldmath ${\lambda}$\unboldmath}}
\newcommand{\n}{\mbox{${\bf n}$}}
\newcommand{\e}{\mbox{${\bf e}$}}
\newcommand{\bi}[1]{\bibitem{#1}}
\newcommand{\fr}[2]{\frac{#1}{#2}}
\newcommand{\p}{\mbox{${\bf p}$}}
\newcommand{\vp}{\mbox{${\bf P}$}}
\newcommand{\q}{\mbox{${\bf q}$}}
\newcommand{\vq}{\mbox{${\bf Q}$}}
\newcommand{\vd}{\mbox{${\bf \Delta}$}}
\newcommand{\f}{\mbox{${\bf f}$}}
\begin{document}

\begin{titlepage}

\begin{center}
{\Large \bf Budker Institute of Nuclear Physics}
\end{center}

\vspace{1cm}

\begin{flushright}
{\bf Budker INP 97-35\\
April 2, 1997 }
\end{flushright}

\vspace{1.0cm}
\begin{center}{\Large \bf High-energy  photon splitting in a strong
Coulomb field.}\\ \vspace{1.0cm}

{\bf R.N.  Lee\footnote{e-mail address: lee@inp.nsk.su},
 A.I. Milstein\footnote{e-mail address: milstein@inp.nsk.su},
 V.M. Strakhovenko\footnote{e-mail address: strakhovenko@inp.nsk.su}}
\\ G.I. Budker Institute of Nuclear Physics,
\\ 630090 Novosibirsk, Russia\\
\end{center}

\bigskip

\begin{abstract}
The helicity amplitudes of the high-energy photon
splitting in the external Coulomb field are obtained
exactly in the parameter $\al$. The cases of screened and unscreened
potentials are investigated. The consideration is based on the
quasiclassical approach, valid for small angles between all photon
momenta. New representation of the quasiclassical electron Green
function is exploited. General expressions obtained are analyzed in
detail for the case of large transverse momenta of both final photons
compared to the electron mass.

\end{abstract}
\end{titlepage}

\section{\bf Introduction}

It is well known, that the virtual electron-positron pair creation
in an external Coulomb field gives rise to such nonlinear QED
phenomena  as Delbr\"uck scattering( coherent photon scattering
\cite{D}) and the splitting of one photon into two. At present
the process of Delbr\"uck scattering has been studied in detail both
theoretically and experimentally (see recent review \cite{MShu}).  At
high photon energies $\omega \gg m$ ($m$ is the electron mass,
$\hbar =c=1$) the scattering amplitude has been found exactly
in the parameter $\al$ ( $Z|e|$ is the nucleus charge,
$\alpha =\, e^2/4\pi\, =1/137$ is the fine-structure constant, $e$ is
 the electron charge). The approaches used essentially depended
on the momentum transfer $\Delta= |\k_2 -\k_1 |$ ($\k_1 $, $\k_2$
being the momenta of the initial and final photons, respectively).
The main contribution to the total cross section of Delbr\"uck
scattering comes from small momentum transfers $\Delta\ll\omega$
(the scattering angle $\theta\sim\,\Delta/\omega\,\ll 1$).
In this case the amplitudes have been found in \cite{CW1,CW2,CW3}
by summing in a definite approximation the diagrams of perturbation
theory with respect to the interaction with the
Coulomb field, and also in \cite{MS1,MS2} within the quasiclassical
approach.  It turned out that at $\omega\gg m$ and $\al\sim 1$ the
exact in  $\al$ result significantly differs from that obtained in
the lowest order of the perturbation theory.

The applicability of the quasiclassical approximation is
based on the fact that, according to the uncertainty
relation, the characteristic impact parameter is $\rho\sim 1/\Delta$
and the corresponding angular momentum is $l\sim
\omega\rho\sim\omega/\Delta \gg 1$ at small scattering angles. This
fact was used in Refs.  \cite{MS1,MS2}, where the quasiclassical
Green function has been derived from the integral representation for
the Green function of the Dirac equation in the Coulomb field
\cite{MS3}.  In Refs.  \cite{LM1,LM2} the quasiclassical Green
function of an electron for an arbitrary decreasing spherically
symmetric potential has been obtained, which allowed one to calculate
the  Delbr\"uck scattering amplitudes in a screened Coulomb
potential.

So far the process of photon splitting has not been observed,
although some events in the experiment performed at DESY \cite{J} were
erroneously interpreted as photon splitting. As it was shown
in \cite{F} these events were due to the electron-positron pair
production accompanied by hard-photon bremsstrahlung. Some
possibilities to observe photon splitting have been discussed in
\cite{MW}.  Photon splitting has been investigated theoretically in
\cite{Sh,CTP,B,JMO,S} in the lowest order of perturbation theory with
respect to the parameter $\al$. The expressions obtained in
\cite{Sh,CTP} are rather cumbersome  and it is difficult to use
them for numerical calculations. Nevertheless, some calculations
based on the results of Ref. \cite{Sh,CTP} have been carried out
in \cite{JMO,S} . Using the Weizs\"acker-Williams method providing the
logarithmic accuracy the amplitudes of the process have been derived
in an essentially simpler form in Ref.  \cite{B}.  The comparison of
the exact cross  section \cite{JMO} with the approximate result
\cite{B} has shown that at high photon energy  the accuracy is better
than $20\%$.  The magnitude of the Coulomb corrections to the lowest
order amplitude of photon splitting has been unknown up to now. At
the present time the experiment dedicated to the observation of
high-energy photon splitting ($\omega\gg m$) in a strong Coulomb
field is held in the Budker Institute  of Nuclear Physics
(Novosibirsk).  Therefore, the theoretical investigation of the
problem is of great interest.

 In the present paper the high-energy photon-splitting  amplitude is
calculated exactly in $\al$ for small angles $f_2$ and $f_3$ between
the momenta $\k_2$ , $\k_3$ of the final photons and the momentum
$\k_1$ of the initial one.  It is the region of small angles that
gives the main contribution to the total cross section of the
process. Besides, small angles and high energies of photons allow one
to use the quasiclassical approach developed in
\cite{MS1,MS2,LM1,LM2} at the consideration of Delbr\"uck scattering.
We consider the case of a pure Coulomb potential as well as the
influence of screening.  The initial representation for the splitting
amplitude is rather cumbersome and contains a thirteen-fold integral.
The quasiclassical approach gives the transparent picture of the
phenomenon and allows one to determine the region of integration
which gives the main contribution to the amplitude.  Without that it
seems impossible to calculate the amplitude.

Our paper is organized as follows. In Sec. II we perform some
transformations of the exact amplitude.  These transformations
essentially simplify further calculations.  Section  III contains the
discussion of the process kinematics. In Sec.  IV the small-angle
approximation for the quasiclassical Green function is derived. In
Sec. V this Green function is applied to the calculation of the
photon-splitting amplitude. In Sec. VI we consider the case of large
transverse momenta of both final photons ($\omega_2f_2\gg m$,
$\omega_3f_3\gg m$). The limiting cases of small momentum transfers
and $\al\ll 1$ for this  amplitude are studied in Sec. VII and Sec.
VIII, respectively. In the last Section the cross sections obtained
in the Born approximation are presented and compared to those
obtained within the Weizs\"acker-Williams method.

\section{\bf Transformation of the amplitude}

According to the Feynman rules the photon-splitting amplitude in the
Furry representation can be written as:

\beq\label{M1}
M=i{e^3}\int\limits d^4x\mbox{Tr}
<x|\hat e_1\mbox{e}^{-ik_1x}{\cal G} \hat e_2^*\mbox{e}^{ik_2x}{\cal G}
\hat e_3^*\mbox{e}^{ik_3x}{\cal G}|x>\,
+\{k_2^{\mu}\leftrightarrow k_3^{\mu},\,
e_2^{\mu} \leftrightarrow e_3^{\mu}\}\, .
\eeq
Here  $e_1^{\mu}$ and $e_{2,3}^{\mu}$ are the polarization vectors of
the initial and final photons, respectively,  $\hat e =
e^{\mu}\gamma_{\mu}=-\e\g,$ $\gamma^{\mu}$ being the Dirac matrices,
${\cal G}=1/(\hat{\cal P}-m+i0)$, and ${\cal P}_{\mu}=i\partial_{\mu}+
g_{\mu 0}(\al /r)$. The matrix element of the operator $\cal G$  is
the Green function of the Dirac equation in the Coulomb field:
$G(x,x\,')=<x|{\cal G}|x\,'>$.

It is convenient to rewrite the expression \eq{M1} in the form,
containing only the Green functions of the 'squared' Dirac equation:

$$D(x,x\,')=<x|{\cal D}|x'>=<x|1/(\hat{\cal P}^2-m^2+i0)|x\,'>$$.

For this purpose we represent the left operator $\cal G$ in \eq{M1}
in the form  ${\cal G}={\cal D}(\hat{\cal P}+m)$ and use the
commutative relation
$$
(\hat{\cal P}+m) \hat
e\,\mbox{e}^{ikx}=\mbox{e}^{ikx}\, [\, -\hat e\, {\cal G}^{-1}+\hat
e\hat k-2\e\p\,\, ].
$$
One can obtain another expression by  similar transformation of the
right operator $\cal G$ in \eq{M1}. Taking a half-sum of these two
expressions and using the identity $\int
dx\,\mbox{Tr}<x|A_1A_2|x>=\int dx\,\mbox{Tr}<x|A_2A_1|x>$, valid for
arbitrary operators $A_1$ and $A_2$, we get:

\beqn\label{M2}
\dst
M=i{e^3}\int\limits d^4x\mbox{Tr}\Biggl\{\Biggl[
\e_1\e_2^*\mbox{e}^{i(k_2-k_1)x}<x|{\cal D}\mbox{e}^{ik_3x}
(\hat e_3^*\hat k_3-2\e_3^*\p){\cal D}|x> +\\
\dst
\e_1\e_3^*\mbox{e}^{i(k_3-k_1)x}<x|{\cal D}\mbox{e}^{ik_2x}
(\hat e_2^*\hat k_2 -2\e_2^*\p){\cal D}|x>+\nonumber\\
\dst
\e_2^*\e_3^*\mbox{e}^{-i(k_2+k_3)x}<x|{\cal D}\mbox{e}^{-ik_1x}
(-\hat e_1\hat k_1 -2\e_1\p){\cal D}|x> \Biggr]\; +\nonumber \\
\dst
\fr{1}{2}\Biggl[<\!x|\mbox{e}^{-ik_1x}(-\hat e_1\hat k_1-2\e_1\p){\cal D}
\mbox{e}^{ik_2x}(\hat e_2^*\hat k_2-2\e_2^*\p){\cal D}
\mbox{e}^{ik_3x}(\hat e_3^*\hat k_3-2\e_3^*\p){\cal D}|x\!>+
(2\leftrightarrow 3)\Biggr]\Biggr\}.\nonumber
\eeqn

The formula \eq{M2} represents the amplitude as a sum
of the terms containing either two or three Green functions:
$M=M^{(2)}+M^{(3)}$.  The amplitude in the form of eq. \eq{M1}
contains terms of different order of magnitude and a strong
compensation occurs.  After taking the trace in eq.
\eq{M2} the expression obtained contains only the terms of the
required order what is the advantage of this formula.

Passing from the time-dependent Green functions to the
energy-dependent ones, taking the integral over time in eq \eq{M2}
and omitting the conventional factor $2\pi\delta
(\omega_1-\omega_2-\omega_3)$, we get for the contribution to $M$
containing three Green functions
\beqn\label{M33} \dst
M^{(3)}=\fr{i}{2}{e^3}\int\fr{d\eps}{2\pi}\int d\r_1d\r_2d\r_3
\exp[i(\k_1\r_1-\k_2\r_2-\k_3\r_3)]\times\\
\dst
\mbox{Tr}\biggl\{[(-\hat e_1\hat k_1-2\e_1\p)D(\r_1,\r_2\,|\eps-\omega_2)]
[(\hat e_2^*\hat k_2-2\e_2^*\p)D(\r_2,\r_3\,|\eps)]\times\nonumber\\
\dst
[(\hat e_3^*\hat k_3-2\e_3^*\p)D(\r_3,\r_1\,|\eps+\omega_3)]\biggr\}
+\,(k_2^{\mu}\leftrightarrow k_3^{\mu}\, ,\,
\e_2\leftrightarrow\e_3)\, .\nonumber
\eeqn
Here $\p=-i\mbox{\boldmath ${\nabla}$\unboldmath}$ differentiates
 the Green function $D$ of the 'squared' Dirac  equation with
respect to its first argument.

Below the Green function $D(\r_1,\r_2\,|\eps)$ will be called the
'electron Green function' for $\eps > 0$ and the 'positron Green
function' for $\eps < 0$. Let the initial photon propagates along $z$
axis.  Then according to the quasiclassical approach developed in
\cite{MS1,MS2,LM1,LM2} the  main contribution to the amplitude at
high energy  arises from the region of integration over the variables
$z_i$ in which $z'\, <\, z$ for the electron Green function
$D(\r,\r\,'|\eps)$ ($\eps>0$ ) and $z'\, >\, z$ for the positron
Green function ($\eps <0$ ). In terms of noncovariant perturbation
theory this range of variables corresponds to the contribution of
intermediate states for which the difference between their energy
$E_n$ and the energy  of the initial state $E_0=\omega_1$ is small
compared to $E_0$. Outside this region at least for one of the
intermediate states $|E_n-E_0|\sim E_0$ and therefore the
 corresponding contribution is suppressed. Besides, there exist
another restriction for the region giving the main contribution to
the amplitude. This restriction follows from the explicit form of the
quasiclassical Green function. It reads $$ z_1<\, z_2 , z_3 \quad
,\quad z_1<0 \quad ,\quad \mbox{max}(z_2,\, z_3)>\,0 \; .  $$ All
these conditions allow one to depict the main contribution to the
 amplitude $M^{(3)}$ in the form of diagrams, shown  in Fig.  1. The
explicit form of vertices is obvious from eq.  \eq{M33}.  The
electron Green functions are marked with left-to-right arrows, and
positron ones with right-to-left arrows. The arrangement of the
diagram vertices is space ordered. With the use of these diagrams one
can easily determine the limits of integration over the energy and
coordinates. The diagrams (a) and (b) correspond to the following
picture: the photon with momentum $\k_1$ produces at the point $\r_1$
a pair of virtual particles which is transformed in the point $\r_2$
into a photon with momentum $\k_2$. Between these two events the
electron (a) or the positron (b) emits a photon with the momentum
$\k_3$ at the point $\r_3$.

Analogously, the expression for the term  $M^{(2)}$ containing two
Green functions reads
\beqn\label{M32}
\dst
M^{(2)}=i{e^3}\int\fr{d\eps}{2\pi}\int d\r_1d\r_2\, \mbox{Tr}
\biggl\{\exp[i(\k_1\r_1-\k_2\r_2-\k_3\r_2)]\, \e_2^*\e_3^*\,
\times  \\
\dst
[(-\hat e_1\hat k_1-2\e_1\p)D(\r_1,\r_2\,|\eps-\omega_1)]
\,D(\r_2,\r_1\,|\eps)]\, + \biggl[
\exp[i(\k_1\r_1-\k_2\r_2-\k_3\r_1)]\, \e_1\e_3^*\,\times \nonumber\\
\dst
D(\r_1,\r_2\,|\eps-\omega_2)]\,
[(\hat e_2^*\hat k_2-2\e_2^*\p)D(\r_2,\r_1\,|\eps)]\, +
\,(k_2^{\mu}\leftrightarrow k_3^{\mu}\, ,\, \e_2\leftrightarrow\e_3)
\,\biggr]\biggr\}\, .\nonumber
\eeqn
The diagrams, corresponding to the representation \eq{M32}, are
shown in Fig. 2.

\section{\bf Kinematics of the process}

According to the uncertainty relation the lifetime of the virtual
electron-positron pair is $\tau\sim |\r_2 -\r_1 | \sim
\omega_1/(m^2+\tilde{\Delta}^2)$, where
$\tilde{\Delta}=\mbox{max}(|\k_{2\perp}|,\,|\k_{3\perp}|)\ll\omega_1$
, $\k_{2\perp}$ and $\k_{3\perp}$ being the transverse components of
the final photon momenta. The characteristic transverse distance
between the virtual particles can be estimated as
$(m^2+\tilde{\Delta}^2)^{-1/2}$, which is much smaller than the
length of the electron-positron loop. The characteristic impact
parameter is $\rho\sim 1/\Delta$ , where ${\bf
\Delta}=\k_2+\k_3-\k_1$ is the momentum transfer. At small
$\k_{2\perp}$ and $\k_{3\perp}$ ($f_{2,3}\ll 1$) we have
\beq\label{Delta}
\vd^2=\, (\k_{2\perp}+\k_{3\perp}\,)^2+\frac{1}{4}
\left(\frac{\k_{2\perp}\,^2}{\omega_2}+
\frac{\k_{3\perp}\,^2}{\omega_3}\right)^2\, .
\eeq
The characteristic angular momentum is $l\sim \omega/\Delta \gg 1$ ,
and the quasiclassical approximation can be applied.

Let us discuss a screened Coulomb potential.
In the Thomas-Fermi model the screening radius is $r_{c}\sim
(m\alpha)^{-1}Z^{-1/3}$ .
If $R\ll 1/\Delta \ll r_{c}$ ($R$ is the nucleus radius), then the
screening is inessential and the amplitude coincides with that in the
pure Coulomb field. At $1/\Delta \sim r_{c}$ the screening
should be taken into account. Obviously, the impact parameters
$\rho\gg r_{c}$ do not contribute to the total cross section. Due to
this fact we shall concentrate ourselves on the momentum transfer
region corresponding to the impact parameter $\rho\leq r_{c}$.
If $\k_{2\perp}\,^2/\omega_2+\k_{3\perp}\,^2/\omega_3 \ll r_{c}^{-1}$
, then it follows from \eq{Delta} that the condition $\rho\leq r_{c}$
holds only when $|{\bf \Delta}_{\perp}|=|\k_{2\perp}+\k_{3\perp}|\geq
r_{c}^{-1}$.  Thus, the main contribution to the amplitude
is given by the region of momentum transfer ${\bf \Delta}_{\perp}$,
restricted from below. In this region $|{\bf \Delta}_{\perp}|\gg
|{\bf \Delta}_{\parallel}|$, that is ${\bf \Delta}\approx {\bf
\Delta}_{\perp}$. It is this region of parameters which we are going
to consider. In addition, at $\omega/(m^2+\tilde{\Delta}^2) \gg r_{c}
$ the angles between $\k_{1,2,3}$ and $\r_{1,2,3}$ are either small
or close to $\pi$ , and corresponding expansions are used in our
calculations.

According to the Furry theorem the photon-splitting amplitude is an
odd function with respect to the parameter $\al$. Due to the
singularity of the Coulomb potential in the momentum space
($-4\pi\al/\Delta^2$) the region of very small momentum transfers
$\Delta \le r_{c}^{-1}$ is essential only in the lowest (linear in
$\al$) order of the perturbation theory . In this order the
Weizs\"acker-Williams method is applicable and the corresponding
large logarithm appears in the cross section integrated over the
angles of one of the final photons \cite{B}.
In next orders of perturbation theory with respect to the parameter
$\al$ the integral over all momenta corresponding to the external
field should be taken provided that their sum is equal to ${\bf
\Delta}$.  Therefore, even at $\Delta \sim r_{c}^{-1}$ each momentum
is not small and the screening can be neglected. In the Born
approximation the screening can be taken into account by multiplying
the amplitude by the factor $[1-F(\Delta^2)]$, where $F(\Delta^2)$ is
the atomic electron form factor. Thus, to find the photon-splitting
amplitude in a screened Coulomb field it is sufficient to solve the
problem in the pure Coulomb field.

\section{\bf Green  function}

Let us pass now to the consideration of the Green function
$D(\r,\r\,'|\eps)$ appearing in eqs. \eq{M33} and \eq{M32}.
In Refs. \cite{LM1,LM2} the representation of this function has been
found in the quasiclassical approximation for an arbitrary
decreasing spherically symmetric potential. For the case of the
Coulomb field, at small  angle $\theta$ between vectors $\r$ and
$-\r\,'$  we obtain from eq. (14) of \cite{LM1}:
\beqn\label{D} \dst
D(\r,\r\,' |\,\eps )= \,\fr{i\mbox{e}^{i\Q(r+r')}}{4\pi\Q rr'}
\int\limits_{0}^{\infty}\,dl l\left[J_{0}(l\theta)+ i\al\fr{(\val
,\n+\n\,')}{l\theta}J_{1}(l\theta)\,\right]\times \\ \dst
\exp{\left[i\fr{l^2(r+r')}{2\Q rr'}\right]}
\left(\fr{4\Q^2rr'}{l^2}\right)^{iZ\alpha\lambda} \quad  ,\nonumber
\eeqn
where $\val = \gamma^0{\bf \gamma}$ , $\Q^2=\eps^2-m^2$ , $\lambda=
\eps/\Q$ , $\n=\r/r$ and $\n\,'=\r\,'/r'$. Taking into account
the relations
$$ \int dl l
J_{0}(l\theta)g(l^2)=\,\fr{1}{2\pi} \int d\q\,\exp({i\q\,\vt})\,
g(q^2) \quad ,\quad \fr{\vt}{\theta}J_{1}(l\theta)=\, -\fr{1}{l}\,
\fr{\partial}{\partial \vt}\, J_{0}(l\theta)\, ,
$$
where $g(l^2)$ is an arbitrary function and $\q$ is a two-dimensional
vector, one can rewrite eq. \eq{D} in the form
\beqn\label{D1}
\dst
D(\r,\r\,' |\,\eps )= \,\fr{i\mbox{e}^{i\Q (r+r')}}{8\pi^2\Q rr'}
\int d\q\, \left[1+\al\fr{\val\q}{q^2}\, \right]\times \\ \dst
\exp{\left [i\fr{q^2(r+r')}{2\Q rr'}+i\q\,(\vt+\vt\,')\right]}
\left(\fr{4\Q^2rr'}{q^2}\right)^{iZ\alpha\lambda} \quad \nonumber .
\eeqn
Here $\vt=\r_{\perp}/r$ , $\vt\,'=\r_{\perp}'/r'$.
Expression \eq{D1} contains only elementary functions and
the angles $\vt$ and $\vt\,'$ appear only in the factor
$\exp[i\q\,(\vt+\vt\,')]$. Therefore, the representation \eq{D1} for
the Green function is very convenient for calculations. At small
angle  between vectors $\r$ and $\r\,'$, for the
case of pure Coulomb field, we get from eq. (15) of
\cite{LM1}:
\beq\label{D2}
D(\r,\r\, ' |\,\eps )=
-\fr{\mbox{e}^{i\Q|\r-\r\,'|}}{4\pi |\r-\r\,'|}\,
\left(\fr{r}{r'}\right)^{iZ\alpha\,\lambda\,\mbox{sign}(r-r')}\, .
\eeq
One can see that in this case the Green function differs from that of
free Dirac equation only by the phase factor. It is easy to
check that after the substitution of the expressions \eq{D1} and
\eq{D2} for the Green functions to the splitting amplitudes \eq{M33}
and \eq{M32} all phase factors of the form $r^{\pm iZ\alpha}$ cancel.
Note that one can replace $\Q$ by $|\eps|-m^2/(2|\eps|)$ in eqs.
\eq{D1} and \eq{D2}. Moreover, one should take the relativistic
correction $m^2/(2|\eps|)$ into account only in the factor
$\exp[i\Q(r+r')]$.

\section{\bf Calculation of the amplitudes $M^{(3)}$ and $M^{(2)}$}

Consider now the diagrams, containing three
Green function (See Fig. 1). Obviously, the contribution of diagram
(b) can be obtained from the contribution of diagram (a) by replacing
$\al\rightarrow -\al$  and changing the overall sign.  It provides
the fulfillment of the Furry theorem: the sum of two contributions
(a) and (b) in Fig. 1 is odd function of $\al$. Therefore, this sum
is equal to the odd in $\al$ part of diagram (a) multiplied by two.
Then, at the calculation of diagram (a) the region of integration
over $z_3$ is divided into two:  $z_3>0$ (photon with the momentum
$\k_3$ is ahead) and $z_3<0$ (photon is behind). In the first
region the angles between vectors $\r_2$ , $\r_3$ and $-\r_1$ are
small. In the second region the angles between $\r_1$, $\r_3$ and
$-\r_2$ are small.  Denote the contribution of the first region to
the diagram (a) as $M_1^{(3)}$ and the contribution of the second
region as $M_2^{(3)}$ .  Introduce vectors
$\f_2=\k_{2\perp}/\omega_2$ and $\f_3=\k_{3\perp}/\omega_3$
($|\f_{2,3}|\ll 1$), as well as $\vt_i=\r_{i\perp}/r_i=\n_{i\perp}$
(i=1,2,3). Taking into account the smallness of the angles we have
$d\r_i\, =r_i^2\, dr_i\, d\vt_i $ .  It is convenient to perform
further calculations in terms of the helicity amplitudes
$M_{\lambda_1\lambda_2\lambda_3}(\k_1,\k_2,\k_3)$.  It is sufficient
to calculate three of them, for instance, $M_{+--}(\k_1,\k_2,\k_3)$ ,
$M_{+++}(\k_1,\k_2,\k_3)$ and $M_{++-}(\k_1,\k_2,\k_3)$ . The others
can be obtained by substitutions. Note that within the accuracy of
our calculations there is no need to take into account
the corrections to the transverse part of the polarization vectors
$\e_{2,3}$. Due to  the relation $\e\,\k=0$ the $z-$component of
the polarization vectors can be expressed via the transverse one
 as $(\e_{2,3})_z=\, -\e_{2,3}\f_{2,3}\,$. So, for a given helicity
one can put the transverse part of the final photon polarization
vector to be equal to the polarization vector of the photon with the
same helicity propagating along $z$ axis. Further the polarization
vector $\e_{+}$ corresponding to the positive helicity is denoted as
$\e$. Then the negative helicity polarization vector $\e_{-}$ is
equal to $\e^{*}$.  Note that to obtain $M_{++-}$ in our approach it
is necessary to calculate both $M_{++-}^{(3)}$ and $M_{+-+}^{(3)}$
since the arrangement of the vertices on the diagrams is space
ordered.

 Let us substitute the expressions \eq{D1} and
\eq{D2} to \eq{M33} and make the obvious expansion at small angles,
taking into account terms, quadratic in $\f_i$ and $\vt_i$.
Introduce notations $\Q_2=\omega_2-\eps\,$ ,
$\Q_3=\omega_3+\eps\,$ and pass to the variables
$$
\q_2\rightarrow\Q_2\q_2\, , \quad \q_3\rightarrow\Q_3\q_3\, , \quad
R_1=\fr{\omega_1}{\Q_2\Q_3}r_1\, , \quad
R_2=\fr{\omega_2}{\eps\Q_2}r_2\, ,\quad R_3=
 \fr{\omega_3}{\eps\Q_3}r_3\, .
$$
Simple integration over $\vt_i$ leads to
\beq\label{MA1}
M_1^{(3)}=\fr{{e^3}}{32\pi^3\omega_1\omega_2\omega_3}
\int\limits_0^{\omega_2}\eps\Q_2\Q_3\, d\eps\int\limits_0^{\infty}dR_1\!
\int\limits_0^{\infty}dR_2\!\int\limits_0^{L}\frac{dR_3}{R_1 R}
\int\!\int d\q_2\,d\q_3 \left(\fr{q_2}{q_3}\right)^{2iZ\alpha}\,
\mbox{e}^{i\Phi}\, T\, ,
\eeq
where $L=R_2\omega_3\Q_2/\omega_2\Q_3$, and
\beqn\label{T}
\dst
T=\fr{1}{4}\mbox{Tr}\biggl\{
\left(1+\fr{\al\,\val\q_3}{\Q_3\,q_3^2}\right)
\left(\fr{2}{R_1}\e_1\,\vq-\hat e_1\hat k_1\right)
\left(1-\fr{\al\,\val\q_2}{\Q_2\,q_2^2}\right)\times \\
\dst
\left[\,\left(\fr{2}{R}(\e_2^*,\vq+\eps R_3\f_{23})
-\hat e_2^*\hat k_2\right)\left(\fr{2}{R}(\e_3^*,\vq+\eps R_2\f_{23})
-\hat e_3^*\hat k_3\right)-\fr{4i}{R}\e_2^*\e_3^*\right]
\biggr\}\, , \nonumber\\
\dst
{}\nonumber\\
\dst
\Phi=\biggl[\left(\fr{1}{R}+\fr{1}{R_1}\right)\fr{\vq^2}{2}
+\fr{\eps^2R_2R_3\,\f_{23}^2}{2R} -
\fr{(\Q_2\q_2-\Q_3\q_3,\vd )}{\omega_1}-\nonumber \\
\dst
\fr{(\omega_3\Q_2R_2-\omega_2\Q_3R_3)}{\omega_1R}(\vq\f_{23})
 -\fr{m^2}{2}(R_1+R)\,\biggr]\, ,\nonumber\\
\dst
R=R_2 -R_3\, ,\quad \f_{23}=\f_2-\f_3\, ,\quad \vq=\q_2+\q_3 \, ,
\quad \vd =\omega_2\f_2+\omega_3\f_3\, . \nonumber
\eeqn
It is convenient to transform the function $T$ in eq. \eq{T} to
the form which does not contain the parameter $\al$. To this purpose
we use the identities
$$
\al\fr{\q_2}{q_2^2}
\left(\fr{q_2}{q_3}\right)^{2iZ\alpha}\,=
-\fr{i}{2}\,\fr{\partial}{\partial\q_2}\left(\fr{q_2}{q_3}\right)^{2iZ\alpha}\,
\quad , \quad
\al\fr{\q_3}{q_3^2} \left(\fr{q_2}{q_3}\right)^{2iZ\alpha}\,=
\fr{i}{2}\,\fr{\partial}{\partial\q_3}
\left(\fr{q_2}{q_3}\right)^{2iZ\alpha}\, ,
$$
and integrate by parts over $\q_2$  and $\q_3$ in \eq{MA1}.
After that some terms in the function $T$ are independent of the
variable $R_1$ and others contain it in the form of factors $1/R_1$,
$1/R_1^2$. Taking the trace and integrating by parts over $R_1$
the terms with the factor $1/R_1^{2}$, we obtain for different
polarizations:
\beqn\label{T1} \dst
T_{+--}=\fr{8}{R_1R^2}(\e\vq)(\e\vq_2)(\e\vq_3)\, ;\\ \dst
{}\nonumber\\
\dst
T_{+++}=-\fr{4}{R_1R^2}\left(\fr{\Q_2}{\Q_3}+\fr{\Q_3}{\Q_2}\right)
(\e\vq)(\e^*\vq_2)(\e^*\vq_3)\, -\fr{\omega_1}{\eps R^3}
\left(\e^*,\, \fr{\omega_2}{\Q_2}\vq_3^2\vq_2 -
 \fr{\omega_3}{\Q_3}\vq_2^2\vq_3\right)\, +\nonumber\\
\dst
 \fr{2i\omega_1^2}{\Q_2\Q_3R^2}(\e^*,\,\vq_2+\vq_3)+
\fr{m^2\omega_1}{\eps R}\left(\e^*,\, \fr{\omega_2}{\Q_2}\vq_2-
 \fr{\omega_3}{\Q_3}\vq_3\right) \, ; \nonumber \\
\dst
{}\nonumber\\
\dst
T_{++-}=-\fr{4}{R_1R^2}\left(\fr{\Q_2}{\eps}+\fr{\eps}{\Q_2}\right)
(\e\vq)(\e\vq_2)(\e^*\vq_3)\, +\fr{\omega_2\omega_3}{\eps\Q_3 R_1 R^2}
(\vq_2^2+\vq_3^2)(\e\vq) - \nonumber \\
\dst
\fr{\omega_1\omega_2}{\Q_2\Q_3 R}\left(\fr{\vq_3^2}{R^2}-m^2\right)
(\e\vq_2)\, +
\fr{4i}{R_1R}\left(\fr{\omega_1\omega_2}{\Q_2\Q_3}-2\right)(\e\vq)+
\fr{2i\omega_1\omega_2}{\Q_2\Q_3R^2}(\e,\vq_2+\vq_3)\, , \nonumber
\eeqn
where $\vq_2=\vq+\eps R_2\f_{23}$ and $\vq_3=\vq+\eps R_3\f_{23}$.
The function $T_{+-+}$ can be obtained from $T_{++-}$ by the
substitution $\omega_2\leftrightarrow\omega_3$ ,
$\Q_2\leftrightarrow\Q_3$ , $\vq_2\leftrightarrow\vq_3$ and $\eps
\rightarrow -\eps $.

In the same way for the contribution $M_2^{(3)}$ we obtain:
\beq\label{MA2}
M_2^{(3)}=\fr{{e^3}}{32\pi^3\omega_1\omega_2\omega_3}
\int\limits_0^{\omega_2}\eps\Q_2\Q_3\, d\eps\int\limits_0^{\infty}dR_1\!
\int\limits_0^{\infty}dR_2\!\int\limits_0^{L_1}\frac{dR_3}{r R_2}
\int\!\int d\q_2\,d\q_3 \left(\fr{q_2}{q_3}\right)^{2iZ\alpha}\,
\mbox{e}^{i\tilde\Phi}\, \tilde T\, ,
\eeq
where $L_1=R_1\omega_3\Q_2/\omega_1\eps$ , $r=R_1+R_3$ ,
\beqn
\dst
\tilde\Phi=\biggl[\left(\fr{1}{r}+\fr{1}{R_2}\right)\fr{\vq^2}{2}
-\fr{\Q_3^2R_1R_3\,\f_{3}^2}{2r} -
\fr{(\Q_2\q_2-\eps\q_3,\vd )}{\omega_2}+\nonumber \\
\dst
\fr{(\omega_3\Q_2R_1-\eps\omega_1R_3)}{\omega_2r}\,(\vq\f_{3})
 -\fr{m^2}{2}(R_2+r)\,\biggr]\, ,
\eeqn
and the function $\tilde T$ for different polarizations reads
\beqn\label{T2}
\dst
\tilde T_{+--}=-\fr{8}{r^2R_2}(\e\vq)(\e\vp_1)(\e\vp_3)\, ;\\
\dst
{}\nonumber\\
\dst
\tilde T_{++-}=\fr{4}{r^2R_2}\left(\fr{\Q_2}{\eps}+\fr{\eps}{\Q_2}\right)
(\e^*\vq)(\e\vp_1)(\e\vp_3)\, +\fr{\omega_2}{\Q_3 r^3}
\left(\e ,\, \fr{\omega_1}{\Q_2}\vp_3^2\vp_1 +
 \fr{\omega_3}{\eps}\vp_1^2\vp_3\right)\, -\nonumber\\
\dst
 \fr{2i\omega_2^2}{\Q_2\eps r^2}(\e,\,\vp_1+\vp_3)-
\fr{m^2\omega_2}{\Q_3 r}\left(\e,\, \fr{\omega_1}{\Q_2}\vp_1+
 \fr{\omega_3}{\eps}\vp_3\right) \, ; \nonumber \\
\dst
{}\nonumber\\
\dst
\tilde T_{+++}=\fr{4}{r^2 R_2}\left(\fr{\Q_2}{\Q_3}+\fr{\Q_3}{\Q_2}\right)
(\e^*\vq)(\e^*\vp_1)(\e\vp_3)\, +\fr{\omega_1\omega_3}{\eps\Q_3 r^2R_2}
(\vp_1^2+\vp_3^2)(\e^*\vq) + \nonumber \\
\dst
\fr{\omega_1\omega_2}{\Q_2\eps r}\left(\fr{\vp_3^2}{r^2}-m^2\right)
(\e^*\vp_1)\, -
\fr{4i}{rR_2}\left(\fr{\omega_1\omega_2}{\Q_2\eps}-2\right)(\e^*\vq)-
\fr{2i\omega_1\omega_2}{\Q_2\eps r^2}(\e^*,\vp_1+\vp_3)\, , \nonumber
\eeqn
where $\vp_1=\vq+\Q_3 R_1\f_{3}$ and  $\vp_3=\vq-\Q_3 R_3\f_{3}$.
One can get the function $\tilde T_{+-+}$ from $\tilde T_{+++}$ by
the substitution $\omega_1\leftrightarrow -\omega_3$ ,
$\Q_2\leftrightarrow -\eps$ , $\vp_1\leftrightarrow\vp_3$ and $\e
\leftrightarrow \e^* $.  Note that the integrand in eqs. \eq{MA2} and
\eq{T2} for helicity amplitudes $M_2^{(3)}$ can be obtained from the
integrand in eqs. \eq{MA1} , \eq{T1} for $M_1^{(3)}$ by means of
substitutions
\beqn\label{subs} \dst \q_{2,3}\rightarrow
-\q_{2,3}\quad  , \quad \omega_1\leftrightarrow \omega_2\quad , \quad
\omega_3\rightarrow -\omega_3\quad ,\quad \Q_3\leftrightarrow\eps
\quad ,
\nonumber\\
\dst
R_1\leftrightarrow R_2\quad ,\quad R_3\rightarrow -R_3\quad ,
\quad \f_{23} \leftrightarrow -\f_3\quad , \quad \f_2\rightarrow
-\f_2\, .
\eeqn
so, that
$$
T_{+--}\rightarrow \tilde T_{+--}\, ,\quad
T_{+-+}\rightarrow \tilde T_{+-+}\, ,\quad
T_{+++}\rightarrow \tilde T_{++-}(\e\leftrightarrow\e^*)\, ,\quad
T_{++-}\rightarrow \tilde T_{+++}(\e\leftrightarrow\e^*)\, .
$$
As to the amplitude $M^{(2)}$, after the integration over the angles
$\vt_i$, we get:
\beqn\label{MM2} \dst M^{(2)}_{+--}=0\quad
,\quad M^{(2)}_{+++}=(\e^*\, ,{\bf M}_{12} +{\bf M}_{13})\quad , \\
\dst
M^{(2)}_{++-}=(\e\, ,{\bf M}_{12} +{\bf M}_{23})\quad , \quad
 M^{(2)}_{+-+}=(\e\, ,{\bf M}_{13} +{\bf M}_{23})\, , \nonumber
\eeqn
where
\beqn\label{MM23}
\dst
{\bf M}_{23}=-\fr{i{e^3}}{16\pi^3}
\int\limits_{-\omega_3}^{\omega_2}\! d\eps
\int\limits_0^{\infty}\!\fr{dR_1}{R_1^2}\!
\int\limits_0^{\infty}\!\fr{dR_2}{R_2^2}
\left[\! R_1+\left(\fr{\Q_2-\Q_3}{\omega_1}\right)^2 R_2\right]
\int\!\!\int d\q_2\,d\q_3 \vq \left(\fr{q_2}{q_3}\right)^{2iZ\alpha}\!
\times \nonumber \\
\dst
\exp\biggl\{i\biggl[\left(\fr{1}{R_1}+\fr{1}{R_2}\right)\fr{\vq^2}{2}
+\fr{\omega_2\omega_3\Q_2\Q_3}{2\omega_1^2}\f_{23}^2R_2 -
\fr{(\Q_2\q_2-\Q_3\q_3,\vd )}{\omega_1}
 -\fr{m^2}{2}(R_1+R_2)\biggr]\biggr\}\, ;\nonumber\\
\dst
{}\nonumber\\
\dst
{\bf M}_{13}=\fr{i{e^3}}{16\pi^3}
\int\limits_{0}^{\omega_2}\! d\eps
\int\limits_0^{\infty}\!\fr{dR_1}{R_1^2}\!
\int\limits_0^{\infty}\!\fr{dR_2}{R_2^2}
\left[\! R_2+\left(\fr{\Q_2-\eps}{\omega_2}\right)^2 R_1\right]
\int\!\!\int d\q_2\,d\q_3 \vq \left(\fr{q_2}{q_3}\right)^{2iZ\alpha}\!
\times \nonumber \\
\dst
\exp\biggl\{i\biggl[\left(\fr{1}{R_1}+\fr{1}{R_2}\right)\fr{\vq^2}{2}
-\fr{\omega_1\omega_3\eps\Q_2}{2\omega_2^2}\f_{3}^2R_1 -
\fr{(\Q_2\q_2-\eps\q_3,\vd )}{\omega_2}
 -\fr{m^2}{2}(R_1+R_2)\biggr]\biggr\}\, ,
\eeqn
and the vector ${\bf M}_{12}$ can be obtained from  ${\bf M}_{13}$
by the substitutions $\omega_2\leftrightarrow \omega_3$ and
$\f_3\leftrightarrow\f_2$.
As we shall see, many terms cancel out in the sum $M^{(2)}+M^{(3)}$.

In general case further transformations of the formulae obtained leads
to four-fold integral with the integrand containing the elementary
functions and this problem will be considered in detail elsewhere.
In what follows we restrict ourselves to the case
$|\k_{2\perp}|=|\omega_2\f_2|\gg m$, $|\k_{3\perp}|=|\omega_3\f_3|\gg
m$ when the amplitudes can be essentially simplified.
This range of the parameters corresponds to large virtuality of the
electron-positron pair  compared to the electron mass, which in this
case can be neglected. Note that the ratio
between the momentum transfer $\Delta=|\vd|$ and the electron mass
$m$ can be arbitrary, since $\Delta$ determines a typical impact
parameter $\rho\sim 1/\Delta$ rather than pair virtuality.

\section{Zero mass limit}

It is easy to see that putting $m=0$ in the expressions obtained
leads to the logarithmic divergences in some terms (in other words,
these terms contain $\log m $ at finite mass). For instance, one
obtains such logarithm in the amplitude $M_1^{(3)}$ integrating over
$R_1$ the terms in $T$ which do not contain the factor $1/R_1$ (see
\eq{T1}). The final result, as it should be, does not contain the
logarithms of mass. But the cancellation of these logarithms between
different terms is rather tricky.

Taking the integral over $R_1$ in \eq{MA1} for $T=T_{+--}$
we do not obtain any logarithm in $M_{+--}^{(3)}$ while
$M_{+--}^{(2)}=0$. Eq. \eq{T1} for $T_{+++}$ and $T_{++-}$ contains
the terms proportional to $\vq_2^2$.  It is convenient to pass in
these terms from the variables $R_2$ and $R_3$ to $R_2$ and
$y=R_3/R$, and then integrate by parts over $y$. In the terms
containing $\vq_3^2$ we pass to the variables $R_3$ and $y=R_3/R$ and
also integrate by parts over $y$.  As a result, in the two-fold
integral over $R_2$ and $R_3$ all logarithms cancel out and one can
put $m=0$.  After that $T_{+++}$ and $T_{++-}$ are transformed to
\beqn\label{T11}
\dst
T_{+++}=-\fr{4}{R_1R^2}\left(\fr{\Q_2}{\Q_3}+\fr{\Q_3}{\Q_2}\right)
(\e\vq)(\e^*\vq_2)(\e^*\vq_3)\quad ;\\
\dst
T_{++-}=-\fr{4}{R_1R^2}\left(\fr{\Q_2}{\eps}+\fr{\eps}{\Q_2}\right)
(\e\vq)[\, (\e\vq_2)(\e^*\vq_3)\, - iR\,]\quad . \nonumber
\eeqn
In addition, there are integrated terms at
$y=\omega_3\Q_2/\omega_1\eps $ ( upper limit) and $y=0$ (lower limit
).  Making the similar transformations for the amplitude $M_2^{(3)}$
we obtain that $\tilde T_{+++}$ and $\tilde T_{++-}$
turn to
\beqn\label{T21} \dst \tilde
T_{++-}=\fr{4}{r^2R_2}\left(\fr{\Q_2}{\eps}+\fr{\eps}{\Q_2}\right)
(\e^*\vq)(\e\vp_1)(\e\vp_3)\quad ;\\
\dst
\tilde T_{+++}=\fr{4}{r^2 R_2}\left(\fr{\Q_2}{\Q_3}+\fr{\Q_3}{\Q_2}\right)
(\e^*\vq)[\, (\e^*\vp_1)(\e\vp_3)\, - iR\,]\quad . \nonumber
\eeqn
The integrated terms corresponding to the lower limit cancel out in
the sum of $M_1^{(3)}$ and $M_2^{(3)}$. Remind that to
calculate the amplitude $M^{(3)}$ we have to find the sum
$M_1^{(3)}+(\k_2\leftrightarrow\k_3\, ,\,\e_2\leftrightarrow\e_3)$ ,
extract the odd in $\al$ part and multiply the result by two. After
that the contribution of the upper-limit integrated terms of
$M_1^{(3)}$ vanishes in the case $M_{+++}$, and in the case
$M_{++-}$ and $M_{+-+}$ gives  the finite result  after the summation
with $(\e{\bf M}_{23})$.  To cancel the logarithmic terms we
exploited the antisymmetry of some integrands with respect to the
substitution $\eps\rightarrow \omega_2-\omega_3-\eps\, ,\,
\q_2\leftrightarrow -\q_3$. The contribution of the upper-limit
integrated terms in $M_2^{(3)}$ vanishes in the case of
$M_{++-}$, and for $M_{+++}$ and $M_{+-+}$ gives the finite result at
$m=0$ after the summation with $(\e^*{\bf M}_{13})$ and $(\e{\bf
M}_{13})$, respectively.  Analogously, the amplitudes $(\e^*{\bf
M}_{12})$ and $(\e{\bf M}_{12})$ cancel singular terms of
$M_2^{(3)}(\k_2\leftrightarrow\k_3\,)$ for the amplitudes $M_{+++}$
and $M_{++-}$. For the amplitude $M_{+-+}$ the upper-limit integrated
terms of $M_2^{(3)}(\k_2\leftrightarrow\k_3\,)$ cancel out.
As a result, the sum of integrated terms and $M^{(2)}$ gives the
additional contributions to helicity amplitudes. We
represent them in the form:
\beqn\label{delM}
\dst
\delta M=-\fr{{e^3}}{4\pi^3} \int\limits_0^{\infty}\!
\fr{dR}{R} \int\!\!\int
\fr{d\q_2\,d\q_3}{\vq^2} \left[\left(\fr{q_2}{q_3}\right)^{2iZ\alpha}
-\left(\fr{q_3}{q_2}\right)^{2iZ\alpha}\right]\, F\quad .
\eeqn
For different polarizations the function $F$  reads
\beqn\label{F} \dst F_{+--}=0\quad ;\quad
F_{+-+}=(\e\vq)\left[
\int\limits_{-\omega_3}^{\omega_2}\!
d\eps\, \fr{\Q_2\Q_3^2}{\omega_1^2\eps}\mbox{e}^{i\psi_1}-
\int\limits_{0}^{\omega_2}\!
d\eps\, \fr{\Q_2\eps^2}{\omega_2^2\Q_3}\mbox{e}^{i\psi_2}\right]\, ;
\nonumber\\
\dst
F_{+++}=(\e^*\vq)\int\limits_{0}^{\omega_2}\!
d\eps\, \fr{\eps\Q_2^2}{\omega_2^2\Q_3}\mbox{e}^{i\psi_2}+
(\omega_2\leftrightarrow\omega_3\, ,\,\f_2\leftrightarrow\f_3)\quad ;
\nonumber\\
\dst
F_{++-}=F_{+-+}(\omega_2\leftrightarrow\omega_3\,
,\,\f_2\leftrightarrow\f_3)\quad ; \\
\dst
{}\nonumber\\
\dst
\psi_1=\fr{\vq^2}{2R}+\fr{\omega_2\omega_3\Q_2\Q_3}
{2\omega_1^2}\f_{23}^2R -
\fr{(\Q_2\q_2-\Q_3\q_3,\vd )}{\omega_1}\quad .\nonumber
\eeqn
The phase $\psi_2$ can be obtained from $\psi_1$ by substitution
\eq{subs}.  To obtain  eq. \eq{delM} we have integrated over one of
the radii.  Note that the singularity of the integrand in \eq{F} at
$\eps=0$ disappears in the total expression for the amplitude of the
process so that no regularization is required.

Putting $m=0$ in phases $\Phi$ and
$\tilde\Phi$, we  take elementary integrals over $R_1$ in \eq{MA1} and
over $R_2$ in \eq{MA2}. After  passing from the variables $q_2$ and
$q_3$ to $\vq=\q_2+\q_3$ and $\q=\q_2-\q_3$, the integral
with respect to $\q$ reads:
\beq
J=\int\fr{d\q}{\vq^2}\,\left(\fr{|\q+\vq|}{|\q-\vq|}\right)^{2iZ\alpha}
\exp(-\fr{i}{2}\q\vd)\quad .
\eeq
Let us change the variable $\q\rightarrow |\vq|\q$ and
represent $\vq$ and $\vd$ as $\vq=|\vq|\vl_1$  and  $\vd=|\vd|\vl_2$,
respectively.  After that it is easy to see that $J$ depends on
$S=(\vl_1\vl_2)$ and $|\vq||\vd|$. Note that in two-dimensional case
the form $P=\epsilon_{ij}\lambda_1^i\lambda_2^j$ is also invariant
with respect to rotations. However, the even powers of $P$ can be
expressed via $S$ ($P^2=1-S^2$), and the odd powers of $P$ change
their sign after reflection. On the other hand, $J$ is invariant
under reflection. It becomes obvious if the reflection of $\q$ is
made along with the reflection of $\vl_1$ and $\vl_2$.  Therefore, $J$
is invariant with respect to the change of variables
$\vq\leftrightarrow\vd$. Therefore, we can represent $J$ as
\beq\label{J}
J=\int\fr{d\q}{\vd^2}\,\left(\fr{|\q+\vd|}{|\q-\vd|}\right)^{2iZ\alpha}
\exp(-\fr{i}{2}\q\vq)\quad .
\eeq
This form is very convenient for further calculations. Now one can
easily take the integrals over $\vq$ and all radii. Summing all
contributions we finally get:
\beqn\label{FINAL} \dst
M=\fr{8{e^3}}{\pi^2\omega_1\omega_2\omega_3\vd^2}
\int{d\q}\,({\bf T}\mbox{\boldmath ${\nabla}$\unboldmath}_{\bf q})\,
{\mbox{Im} }\,\left(\fr{|\q+\vd|}{|\q-\vd|}\right)^{2iZ\alpha}
\quad ;\\
\dst
{}\nonumber\\
\dst
{\bf T}_{+--}=\omega_3\,\e\int\limits_{0}^{\omega_2}\! d\eps\,
\fr{\Q_2^2}{(\e^*\A)}\left[
\fr{(\e\B)\Q_3}{\omega_1{\cal D}_1}
-\fr{(\e\C)\eps}{\omega_2{\cal
D}_3}\right]+
{{\omega_2\leftrightarrow\omega_3}\choose{\f_2\leftrightarrow\f_3}}\quad ;
\nonumber\\
\dst
{}\nonumber\\
\dst
{\bf T}_{+++}=\omega_3\int\limits_{0}^{\omega_2}\! d\eps \,
\Q_2\Biggl[
\e^*\fr{\eps}{\omega_2{\cal D}_3}\left(
\fr{\Q_3-\Q_2}{2}+\fr{(\e^*\f_3)}{(\e^*\A)}(\Q_2^2+\Q_3^2)\right)-
\nonumber  \\
\dst
\e\,\fr{(\e^*\B)(\Q_2^2+\Q_3^2)}{2(\e\A)\omega_1{\cal D}_1}
\Biggr] +\quad
{{\omega_2\leftrightarrow\omega_3}\choose{\f_2\leftrightarrow\f_3}}\quad ;
\nonumber  \\
\dst
{}\nonumber\\
\dst
{\bf T}_{++-}=\omega_3\int\limits_{0}^{\omega_2}\! d\eps \,
\Q_2\Biggl[\e\,\fr{\Q_3}{\omega_1{\cal D}_1}
\left(
\fr{\Q_2-\eps}{2}-\fr{(\e\f_{23})}{(\e\A)}(\Q_2^2+\eps^2)
\right)+\e^*\,\fr{(\e\C)(\Q_2^2+\eps^2)}{2(\e^*\A)\omega_2{\cal
D}_3}\Biggr]+\nonumber  \\
\dst
\omega_2\,\e\int\limits_{-\omega_3}^{0}\! d\eps \,
\Q_3\Biggl[
\fr{\Q_3(\Q_2^2+\eps^2)}{(\e^*\B)}\left(
\fr{(\e^*\f_{23})}{\omega_1{\cal D}_1}+
\fr{(\e^*\f_{2})}{\omega_3{\cal D}_2}
\right)
-\fr{\eps\Q_3+\Q_2^2}{2\omega_1{\cal D}_1} +
\fr{\eps^2-\Q_2\Q_3}{2\omega_3{\cal D}_2}\Biggr]\, .
\nonumber
\eeqn
Obviously, ${\bf T}_{+-+}$ can be obtained from ${\bf T}_{++-}$ by
the substitutions $\omega_2\leftrightarrow\omega_3$,
$\f_2\leftrightarrow\f_3$. In eq. \eq{FINAL} the
following notation is introduced:
\beqn\label{abc} \dst {\cal
D}_1=\fr{\omega_2\Q_3\A^2-\omega_3\Q_2\B^2}{\omega_1\eps}-i0\,\,
,\,\,
{\cal D}_2=\fr{\omega_2\Q_3\tilde{\C}^2-\omega_1\eps\B^2}{\omega_3\Q_2}
\,\, ,\,\,
{\cal D}_3=\fr{\omega_1\eps\A^2+\omega_3\Q_2\C^2}{\omega_2\Q_3}
\,\, ,
\nonumber \\
\dst
\A=\q-\vd+2\Q_2\f_2\quad , \quad \B=\q+\vd-2\Q_3\f_3\quad , \\
\dst
\C=\q+\vd-2\eps\f_{23}\quad ,\quad
\tilde{\C}=\q-\vd+2\eps\f_{23}\quad .\nonumber
\eeqn
At the derivation of \eq{FINAL} we used the identity
$$
\vq\exp(-\fr{i}{2}\q\vq)=
2i\mbox{\boldmath$\nabla$\unboldmath}_{\bf q}\exp(-\fr{i}{2}\q\vq)
$$
and integrated by parts over $\q$. Note that vectors $\e$ and
$\e^*$ appeared in denominators in \eq{FINAL} owing to the
application of the relation $2(\e\A)(\e^*\A)=\A^2$.

\section{Asymptotics at small $\vd$ .}

In the small-angle approximation ($|\f_2|,|\f_3|\ll 1$) the cross
section of the process reads:

\beq \label{cross1}
d\sigma=\fr{\omega_1^2}{2^8 \pi^5}|M|^2\, x(1-x)dx \, d\f_2\, d\f_3\,
\quad,
\eeq
where $x=\omega_2/\omega_1$, so that $\omega_3=\omega_1(1-x)$.
Let us define $\ro=(\omega_2\f_2-\omega_3\f_3)/2$. In terms of the
variables $\ro$ and  $\vd$ the cross section has the form
\beq  \label{cross2}
 d\sigma=|M|^2\,\fr{d\vd\, d\ro\, dx}{2^8 \pi^5 \omega_1^2
\,x(1-x)}\quad,
\eeq
Consider the asymtotics of the amplitudes at $|\vd| \ll |\ro|$.
To this purpose multiply ${\bf T}$ in \eq{FINAL} by
$$
1=\vartheta(q_0^2-\q^2)+\vartheta(\q^2-q_0^2) \quad,
$$
where $|\vd|\ll q_0\ll |\ro|$. Then, for the term in \eq{FINAL}
proportional to $\vartheta(q_0^2-\q^2)$ one can put $\q=0$ and
$\vd=0$ in ${\bf T}$ and integrate by parts over $\q$. After that,
using the relation $\mbox{\boldmath${\nabla}$\unboldmath}_{\bf q}
\vartheta(q_0^2-\q^2)= -2\q\, \delta(q_0^2-\q^2)$ one can easily
take the integral over $\q$ since at $|\q|=q_0\gg |\vd|$ one has
$$
{\mbox{Im} }\,\left(\fr{|\q+\vd|}{|\q-\vd|}\right)^{2iZ\alpha}\approx
4\al\,\fr{\q\vd}{\q^2} \quad .
$$
As a result, in the region $|\q|<q_0$ the term proportional
to $\al$ is independent of $q_0$ and the terms of next orders in
$\al$ are small in the parameter $|\vd|/q_0$.

For the term proportional to $\vartheta(\q^2-q_0^2)$ we get
$$
\mbox{\boldmath ${\nabla}$\unboldmath}_{\bf q}
{\mbox{Im} }\,\left(\fr{|\q+\vd|}{|\q-\vd|}\right)^{2iZ\alpha}\approx
4\al\,\fr{\q^2\vd-2\q(\q\vd)}{|\q|^4}\quad .
$$
We put $\vd=0$ in $\bf T$ and perform  the integration first over
the angles of $\q$ and then over its modulus.
As a result, the main in $q_0/|\ro|$ contribution is independent of
$q_0$ and proportional to \al. Taking  the sum of the contributions
from these two regions and performing the integration over the energy
$\eps$, we get
\beqn\label{zero} \dst
M_{+--}=\fr{4iN(\e\ro)^3}{\ro^4} (\vd\times\ro)_z\quad,\quad
N=\fr{4\al{e^3}\omega_2\omega_3}{\pi\omega_1\vd^2\ro^2}\,;
\\
\dst
{}\nonumber\\
\dst
M_{+++}=N
\Biggl[\e^*\vd+
2(\e\vd)\fr{(\e^*\ro)^2}{\ro^2}\left(1+
\fr{\omega_2-\omega_3}{\omega_1}\ln\fr{\omega_3}{\omega_2}+
\fr{\omega_2^2+\omega_3^2}{2\omega_1^2}(\ln^2\fr{\omega_3}{\omega_2}+\pi^2)\right)
\Biggr]\nonumber\\
\dst
{}\nonumber\\
\dst
M_{++-}=N
\Biggl[\e\vd+2(\e^*\vd)\fr{(\e\ro)^2}{\ro^2}\left(1+
\fr{\omega_1+\omega_3}{\omega_2}(\ln\fr{\omega_3}{\omega_1}+i\pi)+
\fr{\omega_1^2+\omega_3^2}{2\omega_2^2}(\ln^2\fr{\omega_3}{\omega_1}+
2i\pi\ln\fr{\omega_3}{\omega_1})\right)
\Biggr]\quad,\nonumber
\eeqn
where $A_z$ is the projection of the vector $\bf A$
on the direction of $\k_1$.  Substituting \eq{zero} into
\eq{cross2} and performing the elementary integration over the
angles of vectors $\vd$ and $\ro$, we come to the expression
\beq\label{sa} d\sigma=\fr{4Z^2
\alpha^5}{\pi^2}\fr{d\rho^2\,d\Delta^2\,dx}{\rho^4\Delta^2}\,g(x)\quad ,
\eeq
where the function $g(x)$ for different polarizations has the form
\beqn\label{g}
\dst
g_{+--}(x)=x(1-x)\quad ,\\
\dst
g_{+++}(x)=\fr{1}{2}x(1-x)\Biggl[1+
\Biggl(1+(2x-1)\ln\left(\fr{1-x}{x}\right)+\nonumber\\
\dst
\fr{x^2+(1-x)^2}{2}\left(\ln^2\left(\fr{1-x}{x}\right)+\pi^2\right)
\Biggr)^2\,\Biggr]\quad ,
\nonumber\\
\dst
g_{++-}(x)=\fr{1}{2}x(1-x)\Biggl[1+\Biggl|1+
(2/x-1)(\ln(1-x)+i\pi)+\nonumber\\
\dst
\fr{1+(1-x)^2}{2x^2}(\ln^2(1-x)+
2i\pi\ln(1-x))\Biggr|^2\,\Biggr]\quad
,\nonumber\\
\dst
g_{+-+}(x)=g_{++-}(1-x)\quad . \nonumber
\eeqn
Formulae \eq{sa} and \eq{g} are in agreement with the corresponding
results of \cite{B}, obtained in the Weizs\"acker-Williams
approximation.  However, this approach does not allow to obtain the
amplitudes \eq{zero} themselves.  The large logarithm
appears after the integration of \eq{sa} over $\Delta^2$ from
$\Delta_{min}^2$ up to $\rho^2$ where $\Delta_{min}\sim r_c^{-1}$ for
the screened Coulomb potential and $\Delta_{min}\sim
\rho^{2}/\omega_1$ for the pure Coulomb case. It is interesting to
compare the contributions of different helicity amplitudes to the
cross section at $\Delta\rightarrow 0$. In Fig. 3 the function $g(x)$
is shown for different helicities as well as the quantity
\beq\label{gmean}
\bar{g}(x)=g_{+--}(x)+g_{+++}(x)+g_{++-}(x)+g_{+-+}(x)\quad ,
\eeq
which corresponds to the summation over the final photon
polarizations. It is seen that $\bar{g}(x)$ has a wide plateau.

The Coulomb corrections to the photon-splitting amplitude at
$\Delta\rightarrow 0$ are small compared to the Born term \eq{zero}.
One can show that in this case the Coulomb corrections are
proportional to $(\al)^3 \Delta \ln^2(\Delta/\rho)$.
Thus,  they become essential at $\Delta\sim \rho$.
We shall present the detailed investigation of the Coulomb
corrections elsewhere.

\section{Born approximation}

As it was mentioned above, the photon-splitting amplitude obtained
in the lowest Born approximation \cite{Sh,CTP} at arbitrary
energies and momentum transfers is cumbersome and rather difficult to
use.  Therefore, it is interesting to consider the first term of
$\al$ expansion  of \eq{FINAL}. To obtain this asymptotics we perform
the substitution
$$
(\e\mbox{\boldmath ${\nabla}$\unboldmath}_{\bf q})\,
{\mbox{Im} }\,\left(\fr{|\q+\vd|}{|\q-\vd|}\right)^{2iZ\alpha}
\longrightarrow\quad
\al\,\left[\fr{1}{(\e^*,\q+\vd)}-\fr{1}{(\e^*,\q-\vd)}\right]
$$
in \eq{FINAL} and write the quantities ${\cal D}_{1-3}$ in \eq{abc}
as
\begin{eqnarray}\label{newD}
\dst {\cal D}_1=\left(\q+\fr{\Q_2-\Q_3}{\omega_1}\,\vd
\right)^2-\fr{4\omega_2\omega_3\Q_2\Q_3}{\omega_1^2}\f_{23}^2-i0\quad
, \\
\dst
{\cal D}_2=\left(\q+\fr{\Q_2-\epsilon}{\omega_2}\,\vd
\right)^2+\fr{4\omega_1\omega_3\Q_2\epsilon}{\omega_2^2}\f_{3}^2\quad
,\quad
{\cal D}_3=\left(\q-\fr{\Q_3+\epsilon}{\omega_3}\,\vd
\right)^2-\fr{4\omega_1\omega_2\Q_3\epsilon}{\omega_3^2}\f_{2}^2\quad
 .\nonumber
\end{eqnarray}
After that we shift the variable of integration $\q$ in each term so
that the quantities ${\cal D}_{1-3}$ become independent of the angle
$\phi$  of the vector $\q$. For instance, in the terms containing
${\cal D}_1$ we make a substitution $\q\rightarrow
\q-\fr{\Q_2-\Q_3}{\omega_1}\,\vd$. As a result, one can pass to the
variable $z=\exp(i\phi)$ and take easily the contour integral. Taking
the integrals with respect to $|\q|$ and $\epsilon$, we get
for the Born amplitudes
\beqn\label{Born} \dst M_{+--}=\fr{2i\al e^3
(\f_2\times\f_3)_z}{\pi\vd^2(\e^*\f_2) (\e^*\f_3)
(\e^*\f_{23})}\quad,\\ \dst \nonumber\\ \dst \nonumber\\ \dst
M_{+++}=\fr{2(\al) e^3\omega_1}{\pi\vd^2
(\e\f_{23})^2\omega_2\omega_3}\Biggl\{(\e\vd)\Biggl[1+\fr{(\e\f_2)+(\e\f_3)}
{(\e\f_{23})}\ln({a_2\over
a_3})+
\nonumber\\
\dst
\nonumber\\
\dst
\fr{(\e\f_2)^2+(\e\f_3)^2}{(\e\f_{23})^2}\left({\pi^2\over 6}
+{1\over 2} \ln^2({a_2\over
a_3})+\mbox{Li}_2(1-a_2)+\mbox{Li}_2(1-a_3)\right)\Biggr]+\nonumber\\
\dst
\nonumber\\
\dst
{1\over (\e\vd)}\left[\omega_3^2(\e\f_3)^2\fr{a_2}{1-a_2}\left(
1+{a_2\,\ln(a_2)\over
1-a_2}\right)+\omega_2^2(\e\f_2)^2\fr{a_3}{1-a_3}
\left( 1+{a_3\,\ln(a_3)\over 1-a_3}\right)
\right]+\nonumber\\
\dst
\nonumber\\
\dst
\fr{2(\e\f_2)(\e\f_3)}{(\e\f_{23})}\left[\omega_3{a_2\,
\ln(a_2)\over 1-a_2}-
\omega_2{a_3\,\ln(a_3)\over 1-a_3}\right]\Biggr\}\quad ,\nonumber \\
\nonumber\\
\dst
\nonumber\\
\dst
M_{++-}=\fr{2(\al) e^3\omega_2}{\pi\vd^2
(\e^*\f_{3})^2\omega_1\omega_3}\Biggl\{(\e^*\vd)\Biggl[1-
\fr{(\e^*\f_2)+(\e^*\f_{23})}{(\e^*\f_3)}\ln({-a_1\over a_2})
+
\nonumber\\
\dst
\nonumber\\
\dst
\fr{(\e^*\f_2)^2+(\e^*\f_{23})^2}{(\e^*\f_3)^2}\left({\pi^2\over 6}
+{1\over 2} \ln^2({-a_1\over
a_2})+\mbox{Li}_2(1-a_2)+\mbox{Li}_2(1+a_1)\right)\Biggr]+\nonumber\\
\dst
\nonumber\\
\dst
{1\over (\e^*\vd)}\left[\omega_3^2(\e^*\f_{23})^2\fr{a_2}{1-a_2}\left(
1+{a_2\,\ln(a_2)\over
1-a_2}\right)-\omega_1^2(\e^*\f_2)^2\fr{a_1}{1+a_1}
\left( 1-{a_1\,\ln(-a_1)\over 1+a_1}\right)
\right]+\nonumber\\
\dst
\nonumber\\
\dst
\fr{2(\e^*\f_2)(\e^*\f_{23})}{(\e^*\f_3)}\left[
\omega_1{a_1\,\ln(-a_1)\over 1+a_1}
-\omega_3{a_2\,\ln(a_2)\over 1-a_2}\right]\Biggr\}\quad ,\nonumber \\
\nonumber
\eeqn
where
$$
a_1=\fr{\vd^2}{\omega_2\omega_3\f_{23}^2}\quad,\quad
a_2=\fr{\vd^2}{\omega_1\omega_2\f_2^2}\quad,\quad
a_3=\fr{\vd^2}{\omega_1\omega_3\f_3^2}\quad,\quad
\mbox{Li}_2(x)=-\int\limits_0^x\fr{dt}{t}\ln(1-t)\quad.
$$

It follows from \eq{abc}  that $\ln(-a_1)$ should be interpreted
as $\ln(-a_1+i0)=\ln(a_1)+i\pi$. Besides,
$$
\mbox{Li}_2(1+a_1)=\mbox{Li}_2(1+a_1-i0)=\fr{\pi^2}{6}-\ln(1+a_1)
[\ln(a_1)+i\pi]-\mbox{Li}_2(-a_1)
$$
The result \eq{Born} is obtained for $|\vd_\perp|\gg
|\vd_\parallel|$.  One can show that it remains valid in the case
$|\vd_\perp|\sim |\vd_\parallel|$ if the expression \eq{Delta} for
$\vd^2$ is used in \eq{Born}. Actually, in eq. \eq{Born} the
difference between $\vd^2$ and $\vd_\perp^2$ is essential only in the
overall factor $1/\vd^2$.  For a screened Coulomb potential the
amplitudes \eq{Born} should be multiplied by the atomic form factor
$(1-F(\Delta^2))$.  For the case of Moli\`ere potential \cite{M} it
reads
\beq\label{FF}
1-F(\Delta^2)=\Delta^2\sum_{i=1}^{3}\,\fr{\alpha_{i}}{\Delta^2+\beta_{i}^2}
\, ,
\eeq
where
\beqn
\label{coef}
\dst
\alpha_{1}=0.1\quad , \quad \alpha_{2}=0.55 \quad,\quad
\alpha_{3}=0.35 \quad, \quad \beta_{i}=\beta_0 b_i \quad,
\\
\dst \nonumber
\\
\dst \nonumber
b_{1}=6 \quad, \quad b_{2}=1.2 \quad , \quad b_{3}=0.3 \quad ,
\quad \beta_0=\, mZ^{1/3}/121\quad .
\eeqn
Remind that the representation \eq{Born} is valid when
$|\k_{2\perp}|\, , \,|\k_{3\perp}|\gg m$.

\section{Cross section}

As it was suggested in \cite{MW}, to overcome the problems of
background in the measurement of photon splitting one has to register
the events with $|\f_{2,3}|\geq f_0$ where $f_0 \ll 1$ is the angle
determined by the experimental conditions. Let us consider the cross
section integrated over $\f_3$ for $|\f_3|>f_0$. It is interesting to
compare the exact result for this cross section ($d\sigma/dx\,d\f_2$ )
following from \eq{Born} and \eq{cross1} with that obtained in the
Weizs\"acker-Williams approximation ($d\sigma_{approx}/dx\,d\f_2$ ).
The large logarithm corresponds to the contribution of the region
$\Delta\ll \rho=|\omega_2\f_2-\omega_3\f_3)/2|$ where $f_3\approx
xf_2/(1-x)$.  Taking the integral over $\Delta^2$ in eq. \eq{sa} from
$\Delta^2_{min}$ up to $\Delta^2_{eff}$, where (see \cite{B})
$$
\Delta^2_{min}=\Delta^2_{\parallel}=(\omega_1 f_2^2x/2(1-x))^2\quad,
\quad \Delta^2_{eff}=\rho^2=(\omega_1 x f_2)^2\quad ,
$$
and summing over the final photon polarizations
we get for a pure Coulomb potential
\beq\label{wwc}
\fr{d\sigma_{approx}}{dx\,d\f_2}=\fr{8Z^2\alpha^5 }
{\pi^3\omega_1^2}\,\fr{\bar{g}(x)}{x^2f_2^4}\,\ln
\left(\fr{2 (1-x)}{f_2}\right)\,\vartheta(\fr{x}{1-x}f_2-f_0) \quad .
\eeq
For the case of a screened Coulomb potential the approximate cross
section is
\beq\label{wwcs}
\fr{d\sigma_{approx}}{dx\,d\f_2}=\fr{4Z^2\alpha^5 }
{\pi^3\omega_1^2}\,\fr{\bar{g}(x)}{x^2f_2^4}\,\left[2\ln
\left(\fr{\omega_1xf_2}{\beta_0}\right)+\gamma
\right]\,\vartheta(\fr{x}{1-x}f_2-f_0)\quad .
\eeq
The function $\gamma$ in eq. \eq{wwcs} is
\beq\label{gamma}
\gamma=1-\sum\limits_{i=1}^3\,\alpha_i^2 (\ln a_i +1)
-2\sum\limits_{i>j}\,\alpha_i\alpha_j
\fr{a_i\ln a_i-a_j\ln a_j}{a_i-a_j}\quad ,\quad a_i=b_i^2+
\Delta^2_{min}/\beta_0^2
\eeq
and the coefficients $\alpha_i$, $b_i$ and $\beta_0$ are defined in
\eq{coef}. If $\Delta^2_{min}/\beta_0^2\gg 1$ then
$\gamma=-\ln(\Delta^2_{min}/\beta_0^2)$ and eq. \eq{wwcs} turns to
eq. \eq{wwc}.  If $\Delta^2_{min}/\beta_0^2\ll 1$ then
$\gamma=-0.158$.  For the case of a pure Coulomb potential the
dependence of $\sigma_0^{-1} d\sigma/dxd\f_2$ on $f_2/f_0$ is shown
in Fig. 4 at $f_0=10^{-3}$ and $x=0.7$ (curve 1), $x=0.3$ (curve 2),
where
$$ \sigma_0=\fr{4Z^2\alpha^5\bar{g}(x)}{\pi^3\omega_1^2
f_0^4}\quad ,
$$
and  $\bar{g}(x)$ is defined in \eq{gmean}. For $x=0.7$ the cross
section given by \eq{wwc} practically coincides with the curve 1. For
$x=0.3$ it is not the case (the curve 3 corresponds to the cross
section in the Weizs\"acker-Williams approximation at $x=0.3$).
However, within a good accuracy the cross section $d\sigma/dx$ at
$x=0.3$ agrees with that obtained from eq.  \eq{wwc}.  It should be
so, since $d\sigma / dx$ is invariant with respect to the
substitution $x\rightarrow 1-x$ and at $x=0.7$, as we pointed out
above, the approximate result \eq{wwc} is in accordance with the
exact one. At $x=0.5$ the difference between the exact result and the
approximate one is rather essential (see Fig. 5). It can be explained
as follows. The large logarithm appears as a result of
integration with respect to $\f_3$ over the range
$|(1-x)\f_3+x\f_2|\ll x f_2$. After the integration over the
azimuth angle $\varphi$ between vectors $\f_2$ and $-\f_3$ we should
integrate over $f_3$ from $f_0$ up to $x f_2/(1-x)$ and from $x f_2/(1-x)$
to infinity. If $x f_2/(1-x) \approx f_0$ then the contribution
of the first region vanishes and the cross section becomes
approximately two times smaller (in accordance with Fig. 5).

Since the amplitudes \eq{Born} are obtained in zero-mass limit, it is
interesting to estimate the accuracy of this approximation. Numerical
calculations of the Born cross section $d\sigma/dx\,d\f_2$ with the
electron mass taken into account were performed in \cite{JMO} (see
Table V of that paper) for $Z=79$, $x=0.87$, $\omega_1=1.7$ GeV,
$3.4$ GeV, $6.1$ GeV and for five values of the angle $f_2$ in the
interval ($1.2-2.8$) mrad. These results are compared with ours in
Table. One can see a good agreement everywhere. Only in one point
corresponding to the smallest value of the transverse momentum
$k_{2\,\perp}=1.77$ MeV the accuracy is 7\%.

For the case of a pure Coulomb potential the cross section
$d\sigma/dx$ can be approximated with a good accuracy by the
following formula
\beq
\fr{d\sigma_{coul}}{dx} =
\pi f_0^2\,\sigma_0\,\left[\fr{\vartheta(x-1/2)}{x^2}\, \left(2\ln
\fr{2 (1-x)}{f_0}-1-F(x)\right)+(x \leftrightarrow 1-x)\right]\quad ,
\eeq where \beq \label{f}
F(x)=\fr{1}{2}+\fr{x}{1-x}+\fr{2x-1}{(1-x)^2}
\ln\left(2-\fr{1}{x}\right)\quad .
\eeq
If $\omega_1^2 f_0^4/\beta_0^2\ll 1$ then the corresponding
expression for the cross section in a screened Coulomb potential
reads
\beq
\fr{d\sigma_{scr}}{dx} =\pi
f_0^2\,\sigma_0\,\left[\fr{\vartheta(x-1/2)}{x^2}\, \left(2\ln
\fr{\omega_1xf_0}{\beta_0}+0.842-F(x)\right)+(x \leftrightarrow
1-x)\right]\quad ,
\eeq
with $F(x)$ defined in \eq{f}. The function $F(x)$ characterizes the
difference between the exact cross section and that obtained in the
Weizs\"acker-Williams approximation. It is seen from Fig.  6 that
this difference can be significant only at $x$ being close to $0.5$ .
As for the total cross section, the difference between the exact
result and the approximate one is about a few per cent.

The inequality $\Delta\ll \rho$ which provides the applicability
of the Weizs\"acker-Williams approximation corresponds to a small
angle $\varphi$ between the vectors $\f_2$ and $-\f_3$ ( when the
vectors $\f_2$ and  $\f_3$ have almost opposite directions).  So, it
is interesting to consider the quantity $d\sigma(\varphi_{max})/dx$
which is the cross section integrated over the angle $\varphi$ from
$-\varphi_{max}$ to $\varphi_{max}$. In the case of a pure
Coulomb potential the dependence of $(\pi f_0^2\,
\sigma_0)^{-1} d\sigma(\varphi_{max})/dx$ on $\varphi_{max}$ is shown
in Fig.  7 for different $x$ and $f_0=10^{-3}$.  One can see that the
cross section becomes close to its total value at relatively large
$\varphi_{max}$. The same conclusion is also valid for the case of a
screened Coulomb potential.

\newpage \begin{center} Figure captions \end{center}

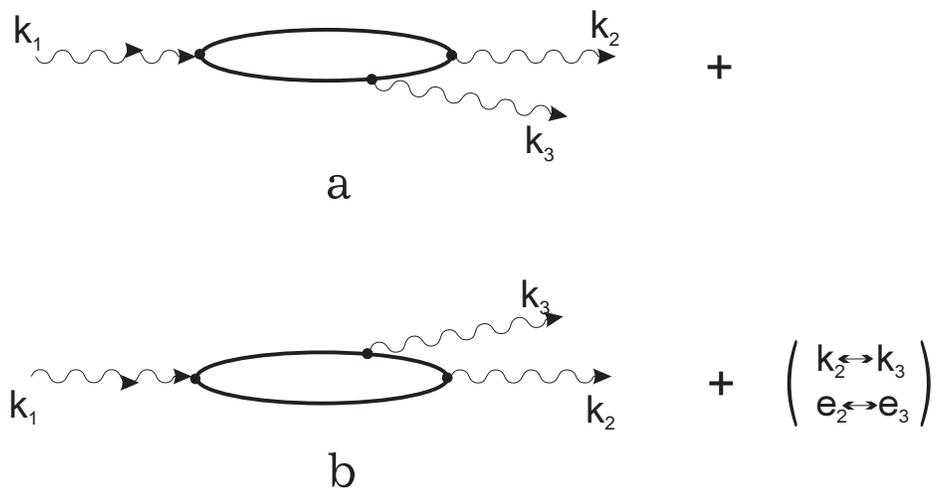
Fig. 1. Diagrams of the perturbation theory
corresponding to the amplitude $M^{(3)}$, eq. \eq{M33}.

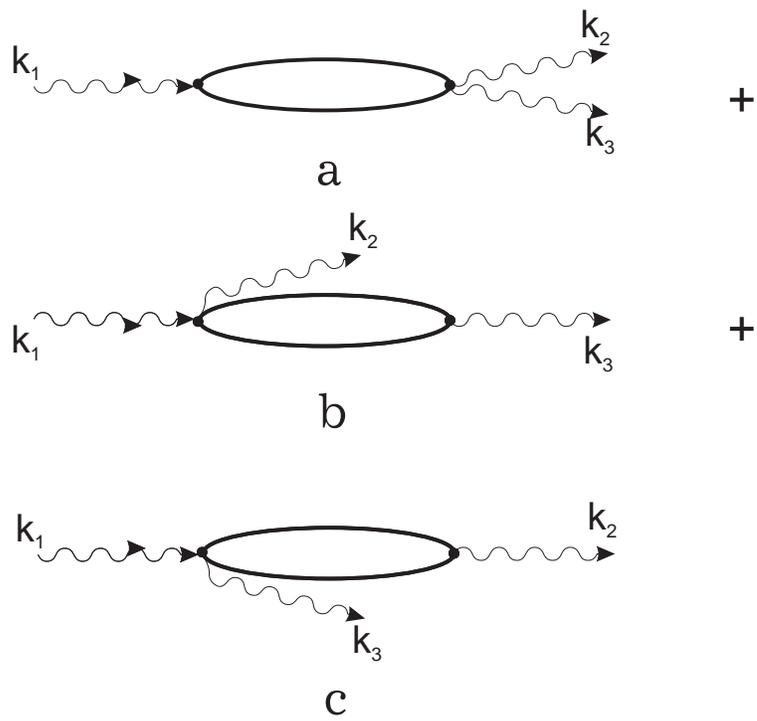
Fig. 2. Diagrams of the perturbation theory
corresponding to the amplitude $M^{(2)}$, eq. \eq{M32}.

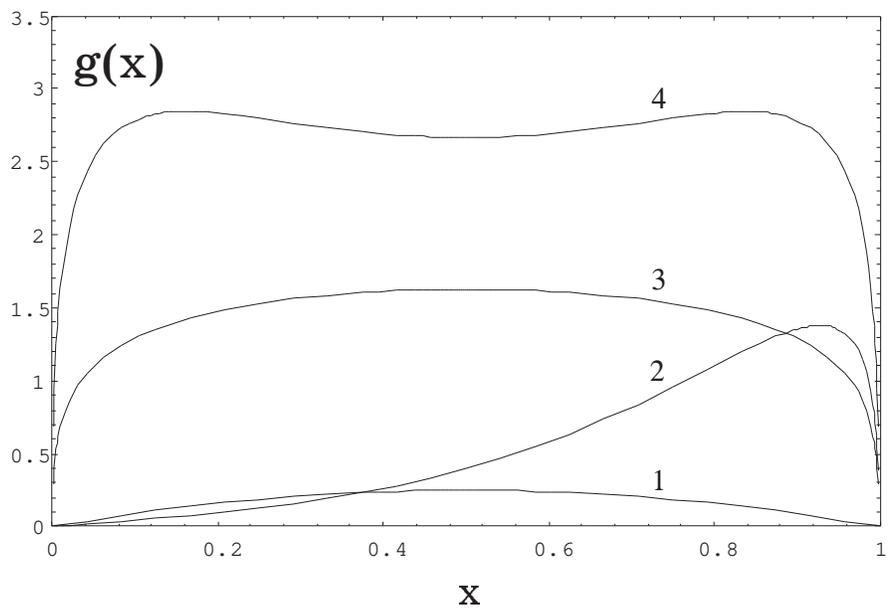
Fig. 3. Function $g(x)$ from eq. \eq{g} for different polarizations:
$g_{+--}(x)$ (curve 1), $g_{++-}(x)$ (curve 2), $g_{+++}(x)$ (curve
3), and $\bar{g}(x)$ (curve 4), eq. \eq{gmean}.

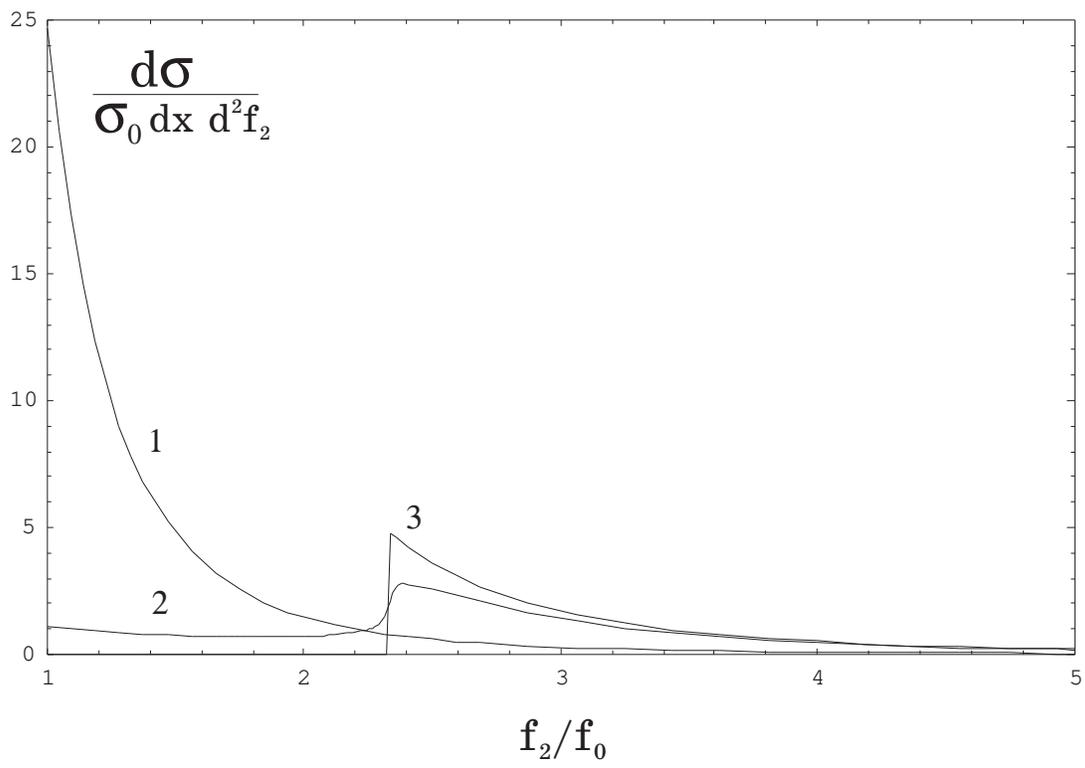
Fig. 4.  $\sigma_0^{-1} d\sigma/dxd\f_2$ versus $f_2/f_0$ for the
case of a pure Coulomb potential, $f_0=10^{-3}$, $x=0.7$ (1), $x=0.3$
(2), $\sigma_0$ is given in the text. Curve 3 corresponds to the
Weizs\"acker-Williams approximation for $x=0.3$.

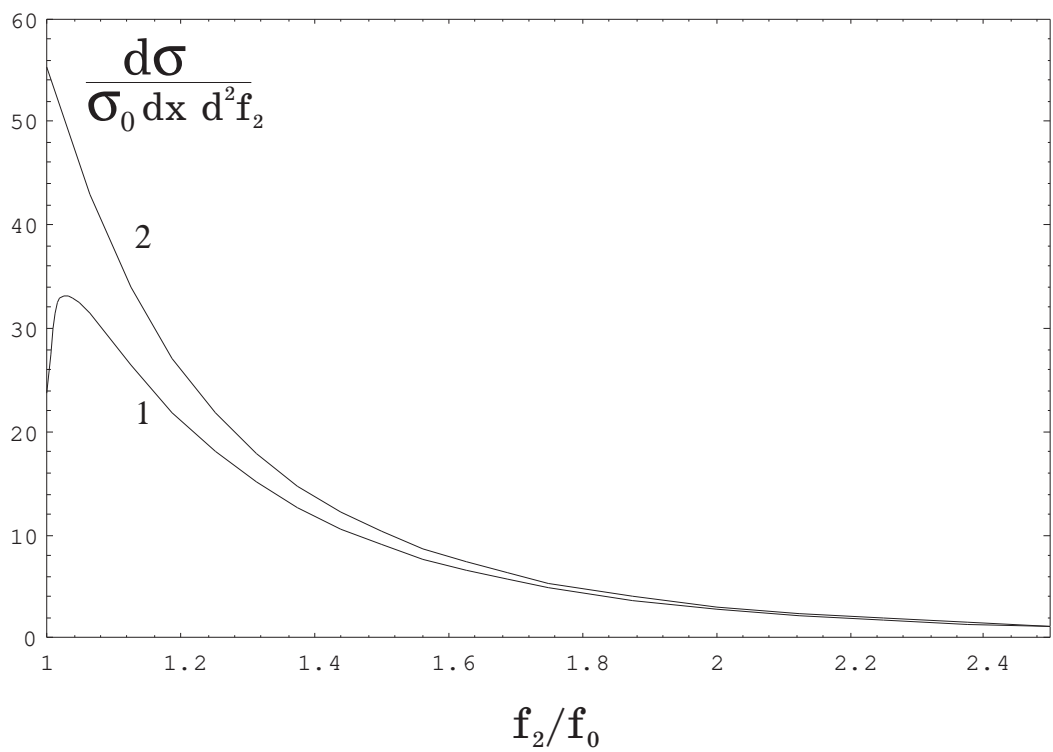
Fig. 5. Same as Fig. 4 but for $x=0.5$ (1). Curve 2 corresponds to
the Weizs\"acker-Williams approximation.

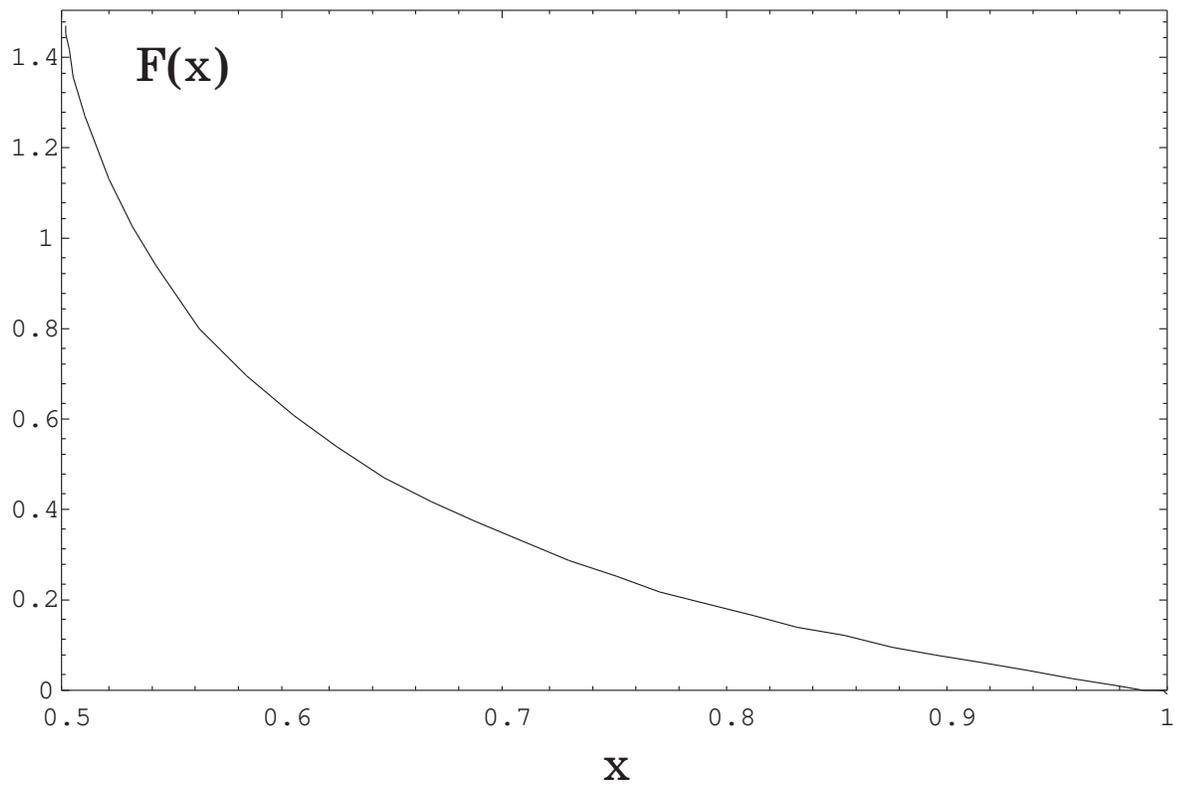
Fig. 6. Function $F(x)$, eq. \eq{f}.

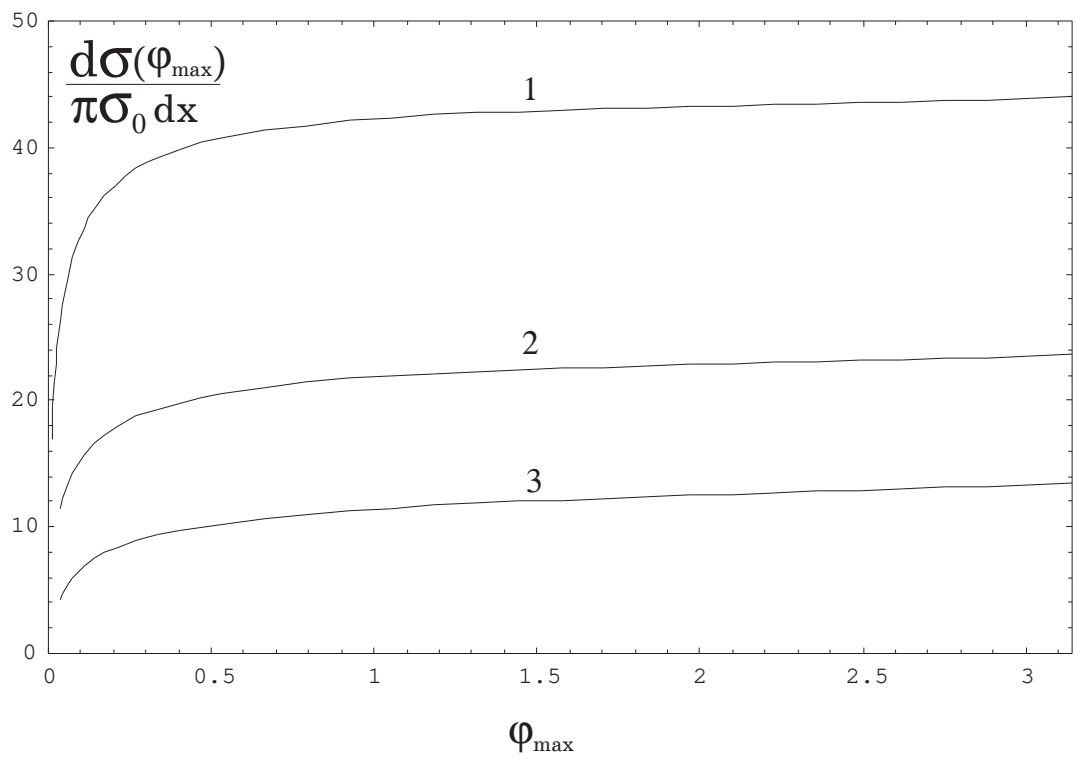
Fig. 7. The dependence of $(\pi f_0^2\,\sigma_0)^{-1}
d\sigma(\varphi_{max})/dx$ on $\varphi_{max}$ for a pure Coulomb case
at different~$x$: $x=0.5$ (1), $x=0.7$ (2), and $x=0.9$ (3);
$f_0=10^{-3}$.

\newpage
Table. {The photon-splitting cross section
$d\sigma/\omega_1 dx\,d\f_2$ in b/GeV sr for $Z=79$, $x=0.87$}

\begin{tabular}{|c|cc|cc|cc|} \hline
~ & \multicolumn{2}{c|}{~} & \multicolumn{2}{c|}{~}
& \multicolumn{2}{c|}{~}\\
$f_2$
& \multicolumn{2}{c|}{ $\omega_1=1.7$ GeV}
& \multicolumn{2}{c|}{ $\omega_1=3.4$ GeV}
& \multicolumn{2}{c|}{ $\omega_1=6.1$ GeV}\\
(mrad) & Present  & Paper \cite{JMO}
& Present  & Paper \cite{JMO}
& Present  & Paper \cite{JMO}\\
~ & result & ~ & result & ~ & result & ~  \\
\hline
1.2 & 22.72 & 21.1 & 3.11 & 3.25 & 0.565 & 0.56 \\
1.6 & 7.47  & 7.4 & 0.994 & 1.03 & 0.177 & 0.18 \\
2.0 & 3.09 & 3.12 & 0.404 & 0.41 & 0.0712 & 0.072 \\
2.4 & 1.48 & 1.51 & 0.191 & 0.19 & 0.0335 & 0.034 \\
2.8 & 0.793 & 0.80 & 0.101 & 0.10 & 0.0177 & 0.018 \\\hline
\end{tabular}

\end{document}